\documentclass{moriond}

\usepackage{siunitx}

\def\be{\begin{equation}}
\def\ee{\end{equation}}
\def\bea{\begin{eqnarray}}
\def\eea{\end{eqnarray}}

\newcommand{\f}{\ensuremath{f_{\text{BBH}}^{\nu}}}

\newcommand{\egw}{\text{3~M}\ensuremath{_{\odot}\ }}

\newcommand*\diff{\mathop{}\!\mathrm{d}}
\newcommand{\figwidth}{.6}

\newcommand{\Photo}{}

\begin{document}
\vspace*{4cm}
\title{IMPLICATIONS OF GW RELATED SEARCHES FOR ICECUBE}



\author{KRIJN D. DE VRIES}
\address{IIHE/ELEM, Vrije Universiteit Brussel, Pleinlaan 2, 1050 Brussels, Belgium}

\author{GWENHA\"EL DE WASSEIGE}
\address{IIHE/ELEM, Vrije Universiteit Brussel, Pleinlaan 2, 1050 Brussels, Belgium}

\author{JEAN-MARIE FR\`ERE}
\address{Universit\'e Libre Bruxelles, Boulevard de la Plaine 2, 1050 Brussels, Belgium}

\author{MATTHIAS VEREECKEN}
\address{IIHE/ELEM, Vrije Universiteit Brussel, Pleinlaan 2, 1050 Brussels, Belgium\\
TENA, Vrije Universiteit Brussel, International Solvay institutes, Pleinlaan 2, 1050 Brussels, Belgium}

\maketitle\abstracts{
At the beginning of 2016, LIGO reported the first ever direct detection of gravitational waves. The measured signal was compatible with the merger of two black holes of about 30 solar masses, releasing about 3 solar masses of energy in gravitational waves.
We consider the possible neutrino emission from a binary black hole merger relative to the energy released in gravitational waves and investigate the constraints coming from the non-detection of counterpart neutrinos, focussing on IceCube and its energy range.
The information from searches for counterpart neutrinos is combined with the diffuse astrophysical neutrino flux in order to put bounds on neutrino emission from binary black hole mergers. Prospects for future LIGO observation runs are shown and compared with model predictions.
}


\section{Introduction}

At the end of 2015, the Laser Interferometer Gravitational-Wave Observatory (LIGO) observed for the first time 
a gravitational wave signal, GW150914.~\cite{Abbott:2016blz} The measured signal is compatible with what is predicted for a binary black hole (BBH) merger. In this event, two black holes with masses 36$^{+5}_{-4}$ M$_ {\odot}$ and 29$^{+4} _{-4}$ M$_ {\odot}$ merged to form a black hole of 62$^{+4}_{-4}$ M$_\odot$, in the process releasing 3$^{+0.5}_{-0.5}$ M$_\odot$ of energy into gravitational waves, from a distance of 410$^{+160}_{-180}$~Mpc from Earth.\\

In the standard picture of binary black hole mergers, such environments are devoid of matter by the time of the merger.~\cite{Perna:2016jqh} Therefore, this type of event is expected to be visible only in gravitational waves.
%
Still, now that binary black hole mergers are observable,~\footnote{Indeed, in addition to GW150914, an additional binary black hole merger has been detected, as well as a candidate event.~\cite{TheLIGOScientific:2016pea}} it is interesting to test whether they do emit other particles or not.
Indeed, the detection of GW150914 triggered several follow-up searches, both in electromagnetic waves~\cite{Abbott:2016gcq} and in neutrinos.~\cite{Adrian-Martinez:2016xgn,Abe:2016jwn,Aab:2016ras} \\

In this work,~\cite{deVries:2016ljw} constraints on high energy neutrino emission from binary black hole mergers are investigated. These constraints come from two sources. Firstly, there are the constraints from direct searches for counterpart neutrinos coincident with the LIGO event. Secondly, given that binary black hole mergers happen throughout the history of the universe, the emission is constrained from the observation of the diffuse astrophysical neutrino flux by IceCube. 

\section{Neutrino emission from BBH mergers}\label{section_f}

The neutrino emission can be characterized by the following parameter
\begin{equation}
\f = \frac{E_\nu}{E_{GW}}.
\label{eq:fractiondef}
\end{equation}
The experimental constraints from follow-up searches for counterpart neutrinos can be immediately translated to a bound on this parameter. \\





Under the assumption that BBH mergers do emit neutrinos, the mergers that have happened throughout the history of the universe give rise to a diffuse flux of neutrinos. According to LIGO,~\cite{TheLIGOScientific:2016pea} the rate of such merger is between $\SIrange{9}{240}{Gpc^{-3} yr^{-1}}$ in the local universe. Following the standard approach,~\cite{Kowalski:2014zda,Ahlers:2014ioa,Waxman:1998yy} one finds that the resulting flux equals
\begin{equation}
E^2 \left.\frac{\diff N_\nu}{\diff E_\nu}\right|_{\mathrm{obs}} = \left( \f t_H \frac{c}{4\pi} \xi_z\right) E^2 \left.\frac{\diff \dot{N}_\nu}{\diff E_\nu}\right|_{\mathrm{inj}, \f=1}.
\label{eq:injtoflux}
\end{equation}
The injection spectrum and rate is contained in $\left.\frac{\diff \dot{N}_\nu}{\diff E_\nu}\right|_{\mathrm{inj}, \f=1}$. Here, it is assumed that the BBH mergers emit neutrinos with an $E^{-2}$-spectrum between \SI{100}{GeV} and \SI{.e8}{GeV}, in order to explain (part of) the observed diffuse astrophysical neutrinos flux. The factor $\xi_z$ captures the cosmic evolution of the sources and the redshift effect on the spectrum. For power law spectra, this $\xi_z$ is $z$-independent. In the following, it will be assumed that BBH mergers follow the star evolution~\cite{Hopkins:2006bw,Yuksel:2008cu} and set $\xi_z = 3$.~\cite{Waxman:1998yy} \\


The resulting flux can then be compared with the diffuse neutrino flux observed by IceCube~\cite{Aartsen:2015zva}
\begin{equation}
E^2 \Phi(E) = 0.84 \pm 0.3\times\SI{.e-8}{GeV cm^{-2} s^{-1} sr^{-1}}.
\label{eq:icdiff}
\end{equation}
One gets a bound on \f\ when the fluxes in Eq.~\ref{eq:injtoflux} and Eq.~\ref{eq:icdiff} are equal. If one would find \f\ higher than this bound from direct observation of BBH mergers, one of the assumptions that went into the calculation must be wrong. This would mean that either \f\ is not universal, or that BBH mergers do not follow star formation, resulting in bounds on these two cases.

%
\section{Prospects}\label{section_prospects}

The bounds on \f\ from the diffuse astrophysical neutrino flux will be compared to those from direct searches as more BBH mergers are accumulated.
All BBH mergers are assumed to be similar to GW150914, emitting \egw of energy in gravitational waves from a distance of 410~Mpc, from random locations in the sky.
The IceCube follow-up analysis of GW150914 is replicated, using the respective effective area,~\cite{Aartsen:2014cva} averaged over the full sky. A background of atmospheric neutrinos~\cite{Sinegovskaya:2013wgm} is included, integrated over a time window of \SI{1000}{s} and over a sky patch of 600~deg$^2$, 100~deg$^2$ and 20~deg$^2$, corresponding to the current~\cite{Abbott:2016blz} and expected~\cite{Gehrels:2015uga} localization uncertainty given by LIGO. \\

\begin{figure}
	\centering
	\includegraphics[width=\figwidth\linewidth]{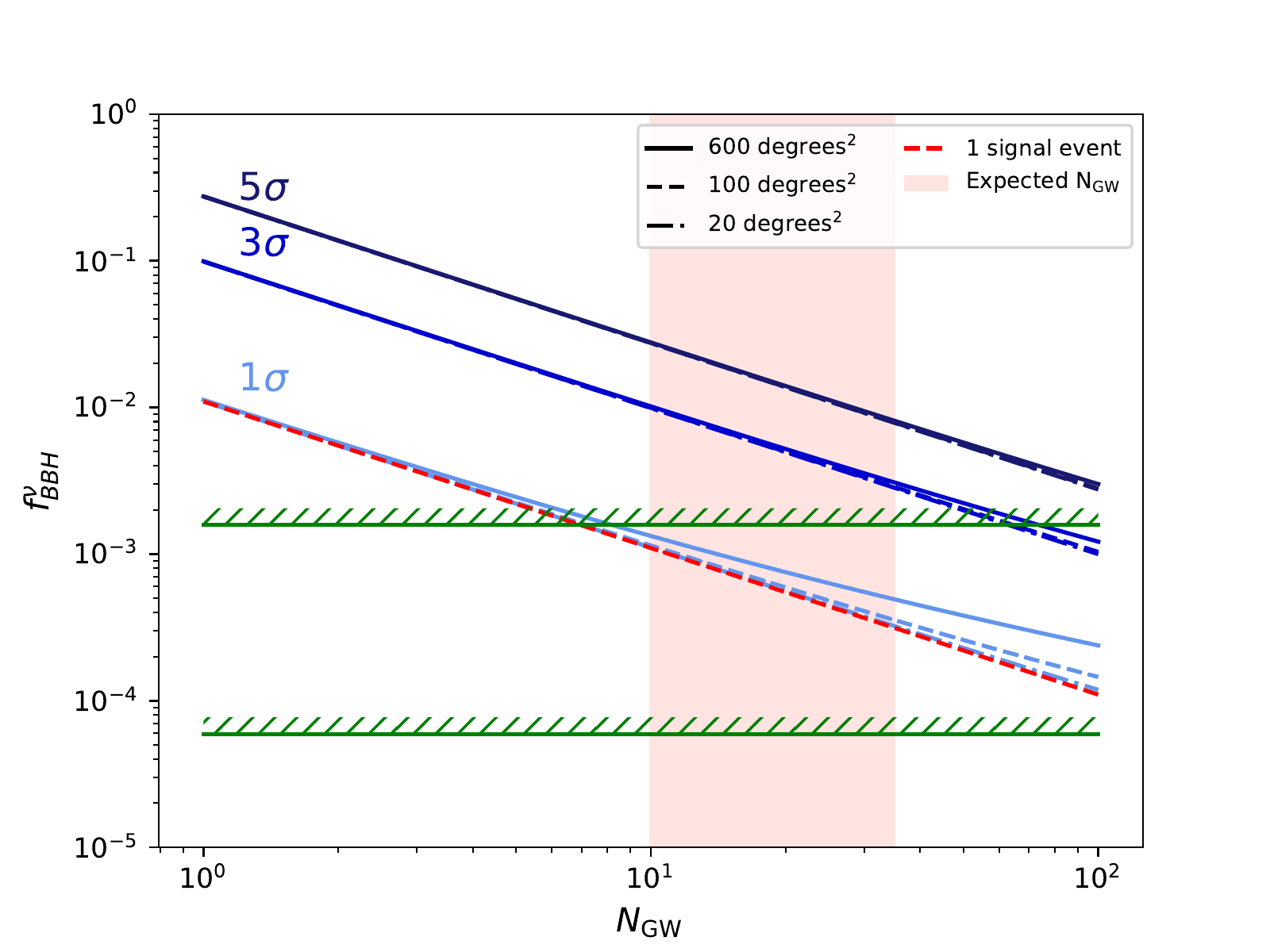}
	\caption{The sensitivity of \f\ expected at various significance requirements (from top to bottom $S/\sqrt{S + B}=5,\,3,\,1$) as a function of number of \egw BBH mergers observed in gravitational waves. The results are shown for different sky patch sizes, corresponding to current and future LIGO pointing accuracy (600~deg$^2$, 100~deg$^2$ and 20~deg$^2$ uncertainty for the full, dashed and dashed-dotted lines respectively).
The green hatched lines show the upper bounds from the astrophysical neutrino flux for the upper and lower limit of the BBH merger rate for the case discussed in the text. The vertical band shows the expected number of BBH mergers seen in LIGO run O2.}
	\label{fig:fracpotential}
\end{figure}

Fig.~\ref{fig:fracpotential} shows the results of this analysis, for various significances with which one could detect a neutrino signal. The astrophysical bounds are at $\f\approx\num{1.58e-03}$ and $\f\approx\num{5.93e-05}$. This value is reached by direct searches, at $S/\sqrt{S + B}=1$, after about 10 mergers have been detected. By the end of LIGO run O2, it is expected that between 10 and 35 BBH mergers (90\% credible interval)~\cite{Abbott:2016nhf} have been seen. In case one would see 35 events, this means that a counterpart neutrino emission could be detected with $S/\sqrt{S + B}=1$ for $\f \approx \num{5e-4}$.

\section{Model dependent interpretation}\label{section_model}

The general bound on \f\ from the previous section can be interpreted in specific models. Assuming now that the neutrino emission originates from matter present around the black hole, \f\ can be decomposed in the following way
\begin{equation}
\f = f_{\mathrm{matter}}\times f_{\mathrm{engine}} \times \epsilon_{\mathrm{p, acc}} \times \epsilon_{\nu}.
\label{eq:fparts}
\end{equation}
In here, $f_{\mathrm{matter}}$ is a fraction expressing the amount of matter present, relative to $E_{\mathrm{GW}}$.
The second part, $f_{\mathrm{engine}} \times \epsilon_{\mathrm{p, acc}}\times \epsilon_{\nu}$ represents the amount of energy that forms an acceleration engine, the amount of energy going into protons and the amount of proton energy going to neutrinos respectively. In the case of a gamma-ray burst fireball model,~\cite{Waxman:1997ti} this last part is equal to $1/10 \times 1/10 \times 1/20$. A bound on \f\ can then be immediately translated into a bound on $f_{\mathrm{matter}}$ within the fireball model.
One can turn this around and estimate possible values of \f, using specific matter models. In the dead disk model,~\cite{Perna:2016jqh} one finds that approximately $\numrange{.e-4}{.e-4}$ M$_\odot$ of matter is present, which can be reactivated upon the merger. This then leads to $\f \approx \num{.e-7}$, still below the reach estimated in Fig.\ref{fig:fracpotential}.

\section{Conclusions}\label{section_concl}

The current and expected future bounds on high energy neutrino emission from binary black hole mergers have been investigated, in terms of a parameter \f.
This was done by combining the information from coincident searches for counterpart neutrinos, together with the already observed diffuse astrophysical neutrino flux. This latter one results in a bound between $\f\approx\numrange{1.58e-03}{5.93e-05}$. It is found that after about 10 binary black hole mergers detected using gravitational waves, it will be possible to probe the neutrino emission down to the upper value of that range. This number of events coincides with the lower value of the 90\% credible interval for expected number of events in LIGO run O2. Finally, when comparing this expected bound with estimates from realistic models of neutrino emission, it is found that such models predict a neutrino emission of $\f \approx \num{.e-7}$, which is below this expected bound in run O2.

\section*{Acknowledgments}

KDV is supported by the Flemish Foundation for Scientific Research (FWO-12L3715N - K.D. de Vries). GDW and JMF are supported by Belgian Science Policy (IAP VII/37) and JMF is also supported in part by IISN. MV is aspirant FWO Vlaanderen. 

\section*{References}

\bibliography{VereeckenMatthias}

\begin{thebibliography}{10}

\bibitem{Abbott:2016blz}
B.~P. Abbott et~al.
\newblock {Observation of Gravitational Waves from a Binary Black Hole Merger}.
\newblock {\em Phys. Rev. Lett.}, 116(6):061102, 2016.

\bibitem{Perna:2016jqh}
Rosalba Perna, Davide Lazzati, and Bruno Giacomazzo.
\newblock {Short Gamma-Ray Bursts from the Merger of Two Black Holes}.
\newblock {\em Astrophys. J.}, 821(1):L18, 2016.

\bibitem{TheLIGOScientific:2016pea}
B.~P. Abbott et~al.
\newblock {Binary Black Hole Mergers in the first Advanced LIGO Observing Run}.
\newblock {\em Phys. Rev.}, X6(4):041015, 2016.

\bibitem{Abbott:2016gcq}
B.~P. Abbott et~al.
\newblock {Localization and broadband follow-up of the gravitational-wave
  transient GW150914}.
\newblock {\em Astrophys. J.}, 826(1):L13, 2016.

\bibitem{Adrian-Martinez:2016xgn}
S.~Adrian-Martinez et~al.
\newblock {High-energy Neutrino follow-up search of Gravitational Wave Event
  GW150914 with ANTARES and IceCube}.
\newblock {\em Phys. Rev.}, D93(12):122010, 2016.

\bibitem{Abe:2016jwn}
K.~Abe et~al.
\newblock {Search for Neutrinos in Super-Kamiokande associated with
  Gravitational Wave Events GW150914 and GW151226}.
\newblock {\em Astrophys. J.}, 830(1):L11, 2016.

\bibitem{Aab:2016ras}
Alexander Aab et~al.
\newblock {Ultrahigh-energy neutrino follow-up of Gravitational Wave events
  GW150914 and GW151226 with the Pierre Auger Observatory}.
\newblock {\em Submitted to: Phys. Rev. D}, 2016.

\bibitem{deVries:2016ljw}
Krijn~D. de~Vries, Gwenhaël de~Wasseige, Jean-Marie Frère, and Matthias
  Vereecken.
\newblock {Constraints and prospects on GW and neutrino emissions using
  GW150914}.
\newblock 2016.

\bibitem{Kowalski:2014zda}
Marek Kowalski.
\newblock {Status of High-Energy Neutrino Astronomy}.
\newblock {\em J. Phys. Conf. Ser.}, 632(1):012039, 2015.

\bibitem{Ahlers:2014ioa}
Markus Ahlers and Francis Halzen.
\newblock {Pinpointing Extragalactic Neutrino Sources in Light of Recent
  IceCube Observations}.
\newblock {\em Phys. Rev.}, D90(4):043005, 2014.

\bibitem{Waxman:1998yy}
Eli Waxman and John~N. Bahcall.
\newblock {High-energy neutrinos from astrophysical sources: An Upper bound}.
\newblock {\em Phys. Rev.}, D59:023002, 1999.

\bibitem{Hopkins:2006bw}
Andrew~M. Hopkins and John~F. Beacom.
\newblock {On the normalisation of the cosmic star formation history}.
\newblock {\em Astrophys. J.}, 651:142--154, 2006.

\bibitem{Yuksel:2008cu}
Hasan Yuksel, Matthew~D. Kistler, John~F. Beacom, and Andrew~M. Hopkins.
\newblock {Revealing the High-Redshift Star Formation Rate with Gamma-Ray
  Bursts}.
\newblock {\em Astrophys. J.}, 683:L5--L8, 2008.

\bibitem{Aartsen:2015zva}
M.~G. Aartsen et~al.
\newblock {The IceCube Neutrino Observatory - Contributions to ICRC 2015 Part
  II: Atmospheric and Astrophysical Diffuse Neutrino Searches of All Flavors}.
\newblock In {\em {Proceedings, 34th International Cosmic Ray Conference (ICRC
  2015): The Hague, The Netherlands, July 30-August 6, 2015}}, 2015.

\bibitem{Aartsen:2014cva}
M.~G. Aartsen et~al.
\newblock {Searches for Extended and Point-like Neutrino Sources with Four
  Years of IceCube Data}.
\newblock {\em Astrophys. J.}, 796(2):109, 2014.

\bibitem{Sinegovskaya:2013wgm}
T.~S. Sinegovskaya, E.~V. Ogorodnikova, and S.~I. Sinegovsky.
\newblock {High-energy fluxes of atmospheric neutrinos}.
\newblock In {\em {Proceedings, 33rd International Cosmic Ray Conference
  (ICRC2013): Rio de Janeiro, Brazil, July 2-9, 2013}}, page 0040, 2013.

\bibitem{Gehrels:2015uga}
Neil Gehrels, John~K. Cannizzo, Jonah Kanner, Mansi~M. Kasliwal, Samaya
  Nissanke, and Leo~P. Singer.
\newblock {Galaxy Strategy for LIGO-Virgo Gravitational Wave Counterpart
  Searches}.
\newblock {\em Astrophys. J.}, 820(2):136, 2016.

\bibitem{Abbott:2016nhf}
B.~P. Abbott et~al.
\newblock {The Rate of Binary Black Hole Mergers Inferred from Advanced LIGO
  Observations Surrounding GW150914}.
\newblock {\em Astrophys. J.}, 833:1, 2016.

\bibitem{Waxman:1997ti}
Eli Waxman and John~N. Bahcall.
\newblock {High-energy neutrinos from cosmological gamma-ray burst fireballs}.
\newblock {\em Phys. Rev. Lett.}, 78:2292--2295, 1997.

\end{thebibliography}

\end{document}